\documentclass[fleqn,twoside]{article}
\usepackage[headings]{espcrc2}

\readRCS
$Id: espcrc2.tex,v 1.2 2004/02/24 11:22:11 spepping Exp $
\usepackage{graphicx}
\usepackage[figuresright]{rotating}

\newcommand{\AmS}{{\protect\the\textfont2
  A\kern-.1667em\lower.5ex\hbox{M}\kern-.125emS}}

\usepackage{amssymb}
\usepackage{amsmath}
\usepackage{color}

\newcommand{\cosths}{\ensuremath{\cos\,\theta^{*}}}
\newcommand{\qcosths}{\ensuremath{q_{\ell}\cosths}}

\title{Single $W$ and Anomalous Single Top Production at HERA}

\author{D.~M.~South\address[MCSD]{Technische Universit\"at Dortmund, Experimentelle Physik V,
        44221 Dortmund, Germany}\thanks{on behalf of the H1 and ZEUS Collaborations.}}

\runtitle{Single $W$ and Anomalous Single Top Production at HERA}
\runauthor{D.~M.~South}

\begin{document}

\begin{abstract}
The search for events containing isolated leptons (electrons or muons)
and missing transverse momentum produced in $e^\pm p$ collisions is
performed individually and in a common phase space with the H1 and ZEUS
detectors at HERA in the period 1994--2007.
The presented H1+ZEUS data sample corresponds to an integrated luminosity of
0.97~fb$^{-1}$, and comprises the complete high energy data from the
HERA programme.
A total of 87 events are observed in the data, compared to a
Standard Model prediction of 92.7~$\pm$~11.2.
At large hadronic transverse momentum $P_{T}^{X} >$~25~GeV in
the $e^{+}p$ data, luminosity 0.58~fb$^{-1}$, 23 data events are
observed compared to a SM prediction of 14.6~$\pm$~1.9.
Production cross section measurements of events containing isolated
leptons and missing transverse momentum and of single $W$ production are
performed by H1, where the measured cross sections are found to be in
agreement with SM predictions.
A complementary search by H1 for events containing an isolated tau lepton and
missing $P_{T}$ is also presented.
A measurement of the $W$ polarisation fractions is performed by H1,
where the presented results are found to be in agreement with the SM.
Finally, the H1 isolated lepton events are examined in the context of a
search for anomalous single top production.
In the absence of a clear signal, an upper limit on the anomalous top
production cross section $\sigma_{ep\rightarrow etX} < 0.16$~pb is established
at the $95\%$ confidence level, corresponding to a limit an upper bound on the
anomalous magnetic coupling $\kappa_{tu\gamma} < 0.14$.
\vspace{1pc}
\end{abstract}

\maketitle


\section{The HERA Electron--Proton Collider}
\label{sec:intro}

\vspace{-1cm}
\begin{figure}[h]
\includegraphics[width=.47\textwidth]{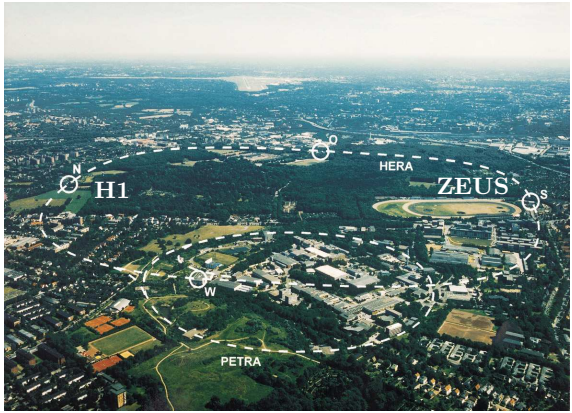}
\vspace{-1cm}
\begin{picture} (0.,0.)
\setlength{\unitlength}{1.0cm}
\put (5.75,3.3){\small \bf \color{white} ZEUS}
\put (1.20,3.25){\small \bf \color{white} H1}
\end{picture}
\caption{The HERA $e^{\pm}p$ collider at DESY. The location of
	the H1 and ZEUS experiments is indicated, as well as the
	pre-accelerator PETRA.}
\label{fig:hera}
\vspace{-0.5cm}
\end{figure}

The HERA $e^{\pm}p$ collider, located in Hamburg, Germany, and shown in
figure \ref{fig:hera}, was in operation in the years 1992--2007. Protons with
an energy up to 920 GeV were brought into collision with electrons or positrons
of energy 27.6 GeV at two experiments, H1 and ZEUS, each of which collected
about 0.5~fb$^{-1}$ of data.
Together with measuring the structure of the proton, the deep inelastic collisions
(DIS) produced at HERA, at a centre of mass energy up to 318~GeV, provided an ideal
environment to study rare processes, set constraints on the Standard Model (SM)
and search for new particles and physics beyond the Standard Model (BSM).


\section{Events with Isolated Leptons and $P_{T}^{\rm miss}$}
\label{sec:isointro}

Events containing a high transverse momentum ($P_{T}$) isolated electron or muon
and missing $P_{T}$ have been observed at
HERA~\cite{isoleph1origwpaper,isoleph1newwpaper,isolepzeusorigwpaper,zeustop}.
An excess of HERA~I data events (1994--2000 which is mostly in $e^{+}p$ collisions)
compared to the SM prediction at large hadronic transverse
momentum $P_{T}^{X}$ was reported by the H1 Collaboration~\cite{isoleph1newwpaper}.
This was not confirmed by the ZEUS Collaboration, although the analysis was
performed in a more restricted phase space~\cite{zeustop}.

\begin{figure}[t]
\includegraphics[width=.47\textwidth]{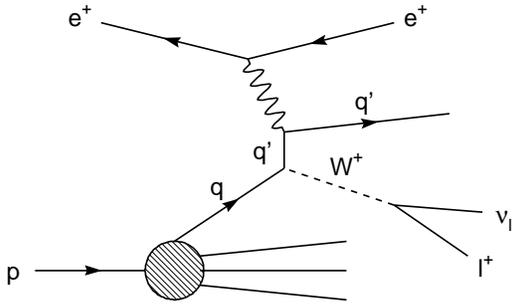}
\vspace{-1cm}
\caption{Feynman diagram of the process
	$ep \rightarrow eW^{\pm}$($\rightarrow l\nu$)$X$, which is the main
	SM signal contribution to the search for events with isolated leptons
	and missing transverse momentum.}
\label{fig:feyn}
\end{figure}

The main SM contribution to the signal topology is the production
of real $W$ bosons via photoproduction with subsequent leptonic decay
$ep\rightarrow eW^{\pm}$($\rightarrow l\nu$)$X$, as illustrated in
figure~\ref{fig:feyn}, where the hadronic system $X$ has typically low $P_{T}$.
The equivalent charged current (CC) process,
$ep \rightarrow \nu$$W^{\pm}$($\rightarrow l\nu$)$X$, also contributes to the
total signal rate, although only at a level of about 7\%.
The production of $Z^{0}$ bosons with subsequent decay to neutrinos
$ep \rightarrow eZ^{0}$($\rightarrow \nu\bar{\nu}$)$X$ results in a
further minor contribution\footnote{This process is not included in
the present ZEUS analysis.} to the total signal rate in the electron
channel at a level of 3\%.
SM signal processes are modelled using the event generator EPVEC~\cite{epvec}.
The main $W$ production via photoproduction component is reweighted to a NLO
calculation~\cite{nlo}, which has a theoretical uncertainty of 15\%.


SM background enters the electron channel due to mismeasured neutral
current (NC) events and the muon channel due to lepton pair (LP)
events in which one muon escapes detection, both cases resulting in
apparent (fake) missing transverse momentum.
CC background, which contains intrinsic missing transverse momentum,
enters the final sample in both lepton channels, where a final
state particle is interpreted as the isolated electron or muon.


The event selection employed by the H1~\cite{h1isolepnew} and
ZEUS~\cite{zeusisolepnew} analyses is very similar and may be summarised
as follows:
The identified lepton should have high transverse momentum $P_{T}^{l} >$~10~GeV,
be observed in the central region of the detector and be isolated with
respect to jets and other good quality tracks in the event.
The event should also contain a large transverse momentum imbalance,
$P_{T}^{\rm miss} >$~12~GeV. Further cuts are then applied, which are designed
to reduce SM background, whilst preserving a high level of signal purity.
Event quantities which are sensitive to the presence of undetected energetic
particles in the event are employed such as the azimuthal balance of the
event, the difference in azimuthal angle between the lepton and the
hadronic system and the longitudinal momentum imbalance.
To ensure that the two lepton channels are exclusive, and may therefore
be combined, electron events must contain no isolated muons.


\subsection{Results from H1 and ZEUS Analyses}
\label{sec:sep}

Both H1 and ZEUS have recently performed the analysis of the electron and
muon channels on their respective complete HERA I+II $e^{\pm}p$ data sets,
which correspond to approximately 0.5~fb$^{-1}$ per
experiment~\cite{h1isolepnew,zeusisolepnew}.


A total of 59 events are observed in the H1 data, compared to a
SM prediction of 58.9~$\pm$~8.2.
For $P_{T}^{X} >$ 25~GeV, a total of 24 events are observed compared
to a SM prediction of 15.8~$\pm$~2.5, where 21 events are observed in
the $e^{+}p$ data compared to a SM prediction of 8.9~$\pm$~1.5.
The observed data excess in the HERA~I $e^{+}p$ data~\cite{isoleph1newwpaper}
thus remains at the 3.0$\sigma$ level for the complete H1 $e^{+}p$
data~\cite{h1isolepnew}.


In the ZEUS analysis of the complete HERA~I+II data, 41 data events are
observed in agreement with the SM prediction of 48.3~$\pm$~6.8~\cite{zeusisolepnew}.
Unlike in the H1 analysis, agreement between data and SM is also observed
in the high $P_{T}^{X}$ region, where 11 events are seen in
the $e^{\pm}p$ data compared to a SM prediction of 13.1~$\pm$~1.8.


\subsection{A Combined H1 and ZEUS Analysis}
\label{sec:comb}

A study of the selection efficiency for signal processes found the
H1 and ZEUS analyses to be compatible in the kinematic region where they
are directly comparable~\cite{h1ichep06,zeusichep06}.
The majority of the data events observed by H1 at $P_{T}^{X} >$ 25~GeV
are also found to be in the region of overlap of the two analyses.
Nevertheless, in order to coherently combine the results from the two
experiments, a common phase space has been established.


This common selection is based\footnote{An additional change with respect
to the H1 event selection is that the cut on longitudinal momentum imbalance
in the electron channel is simplified in that it is always applied. This is
found to make only a negligible difference to the SM expectation.}
on the H1 event selection~\cite{isoleph1newwpaper,h1isolepnew}, but covers a
more restricted lepton polar angle range of
15$^\circ~<~\theta_l~<$~120$^\circ$, as that employed in the ZEUS
analysis~\cite{zeusisolepnew}.
The signal expectation rates of the H1 and ZEUS analyses using the common
selection are found to be comparable, taking into account the
respective luminosities of the data sets and the signal processes included.
More details on the combination of the H1 and ZEUS analyses can be found
in~\cite{h1andzeusisolepnew}.


\begin{table*}[t]
\begin{center}
\caption{Summary of the combined H1+ZEUS results in the search for
	events with isolated electrons or muons and missing transverse
	momentum, shown for the electron and muon channels
	separately and combined for the full HERA~I+II $e^{+}p$,
	$e^{-}p$ and $e^{\pm}p$ data. The number of observed data events is
	compared to the SM prediction. The results are shown for the full
	sample and for events with $P_{T}^{X}>25$~GeV.
	The signal component of the SM expectation, dominated by real $W$
	production, is given as a percentage of the total SM prediction in parentheses.
	The uncertainties contain statistical and systematic uncertainties added
	in quadrature.}
\vspace{0.2cm}
\label{tab:h1zeus}
  \begin{tabular}{ccccc}
    \hline
    \multicolumn{2}{c}{H1+ZEUS Preliminary} &
    Electron &
    Muon &
    Combined \\
    \multicolumn{2}{c}{$l$+$P_{T}^{\rm miss}$ events at} &
    obs./exp. &
    obs./exp. &
    obs./exp. \\
    \multicolumn{2}{c}{HERA I+II} &
    {\small (Signal contribution)} &
    {\small (Signal contribution)} &
    {\small (Signal contribution)} \\
    \hline
    {\small 1994-2007 $e^{+} p$} &
    {\small Full Sample} &
    {\small 39 / 41.3 $\pm$ 5.0 (70\%)}&
    {\small 18 / 11.8 $\pm$ 1.6 (85\%)}&
    {\small 57 / 53.1 $\pm$ 6.4 (73\%)}\\
    \cline{2-5}
    {\small 0.58 fb$^{-1}$} &
    {\small $P_{T}^{X}~>25$~GeV} &
    {\small 12 /  7.4 $\pm$ 1.0 (78\%)}&
    {\small 11 /  7.2 $\pm$ 1.0 (85\%)}& 
    {\small 23 / 14.6 $\pm$ 1.9 (81\%)}\\
    \hline
    {\small 1998-2006 $e^{-} p$} &
    {\small Full Sample} &
    {\small 25 / 31.6 $\pm$ 4.1 (63\%)}&
    {\small  5 /  8.0 $\pm$ 1.1 (86\%)}&
    {\small 30 / 39.6 $\pm$ 5.0 (68\%)}\\
    \cline{2-5}
    {\small 0.39 fb$^{-1}$} &
    {\small $P_{T}^{X}~>25$~GeV} &
    {\small  4 /  6.0 $\pm$ 0.8 (67\%)}&
    {\small  2 /  4.8 $\pm$ 0.7 (87\%)}& 
    {\small  6 / 10.6 $\pm$ 1.4 (76\%)}\\
    \hline
    {\small 1994-2007 $e^{\pm} p$} &
    {\small Full Sample} &
    {\small 64 / 72.9 $\pm$  8.9 (67\%)}&
    {\small 23 / 19.9 $\pm$  2.6 (85\%)}&
    {\small 87 / 92.7 $\pm$ 11.2 (71\%)}\\
    \cline{2-5}
    {\small 0.97 fb$^{-1}$} &
    {\small $P_{T}^{X}~>25$~GeV} &
    {\small 16 / 13.3 $\pm$ 1.7 (73\%)}&
    {\small 13 / 12.0 $\pm$ 1.6 (86\%)}& 
    {\small 29 / 25.3 $\pm$ 3.2 (79\%)}\\
    \hline
  \end{tabular}
\end{center}
\end{table*}

\begin{figure}[tp]
  \includegraphics[width=.47\textwidth]{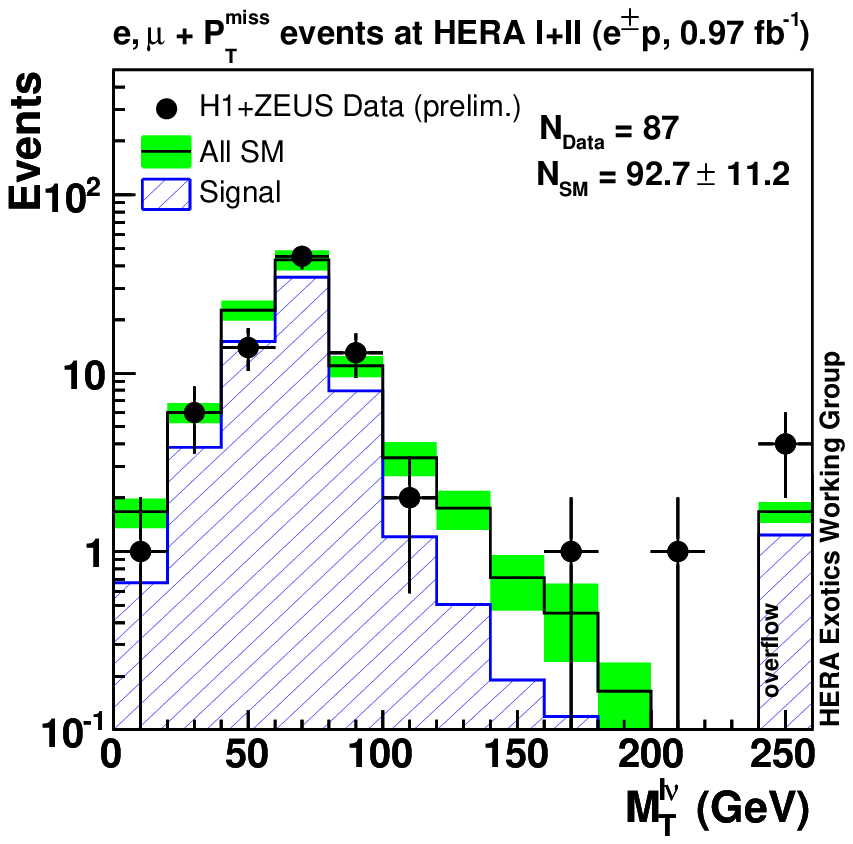}
  \includegraphics[width=.47\textwidth]{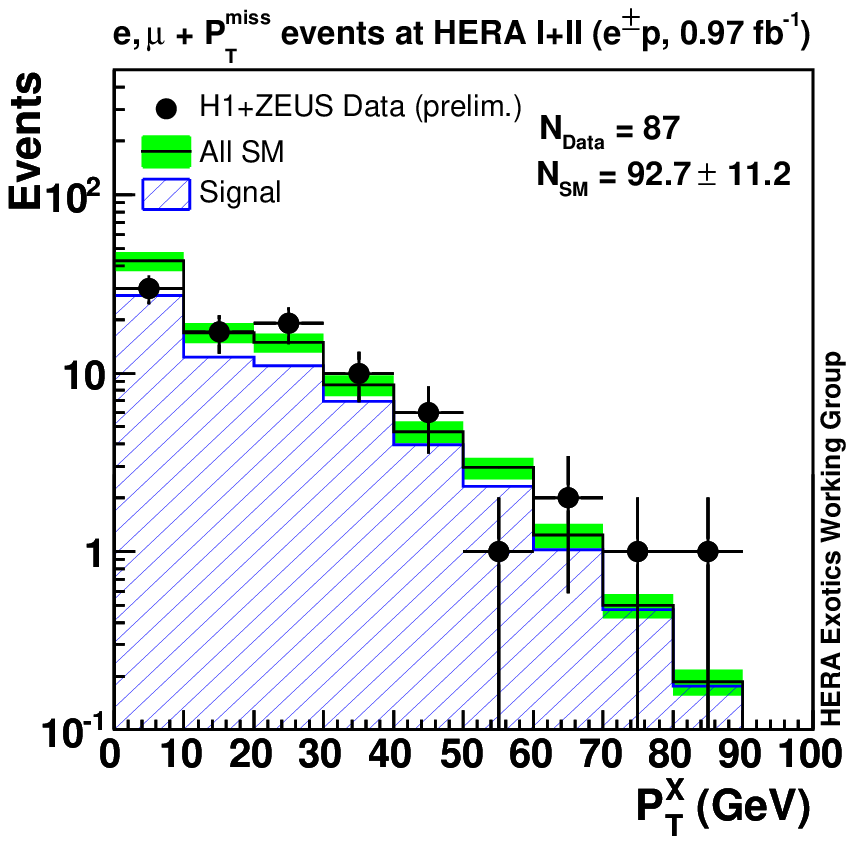}
  \caption{The transverse mass $M_{T}^{l\nu}$ (top) and hadronic
	transverse momentum $P_{T}^{X}$ (bottom)
	distributions of the combined H1+ZEUS $e^{\pm}p$ HERA I+II data.
	The data (points) are compared to the
	SM expectation (open histogram). The signal component of the SM
	expectation is given by the striped histogram. $\rm N_{Data}$ is the
	total number of data events observed and $\rm N_{SM}$ is the total SM
	expectation. The total uncertainty on the SM expectation is given by the
	shaded band.}
\label{fig:h1zeus}
\end{figure}

The results of the combined H1+ZEUS analysis are summarised in
table \ref{tab:h1zeus}.
The signal contribution, mainly from real $W$ production, is seen to
dominate the total SM expectation in all data samples.
At large hadronic transverse momentum $P_{T}^{X} >$ 25~GeV a total of
29 events are observed in the H1+ZEUS $e^{\pm}p$ data compared to a
SM prediction of 25.3~$\pm$~3.2.
In the $e^{+}p$ data alone, 23 events are observed with $P_{T}^{X} >$ 25~GeV
compared to a SM prediction of 14.6~$\pm$~1.9, which is equivalent to an
excess of data over the SM prediction of 1.8$\sigma$.
Seventeen of the 23 data events are observed in the H1 data compared to
a SM expectation of 7.1~$\pm$~0.9, equivalent to an excess of data over
the SM prediction of 2.9$\sigma$.


Figure~\ref{fig:h1zeus} (top) shows the transverse mass, $M_{T}^{l\nu}$
and $P_{T}^{X}$ distributions of the H1+ZEUS $e^{\pm}p$ HERA I+II data for the
combined electron and muon channels.
The distribution of events in $M_{T}^{l\nu}$ is compatible with the
Jacobian peak expected from $W$ production.
Similarly, the observed $P_{T}^{X}$ spectrum displayed in figure~\ref{fig:h1zeus}
(bottom) is compatible with that expected from $W$ production, peaking at low
values of hadronic transverse momentum.


\section{Cross Sections}
\label{sec:xs}

The H1 results described in section \ref{sec:sep} are used to
calculate production cross sections for events with an energetic isolated
lepton and missing transverse momentum ($\sigma_{\ensuremath{\ell+{P}_{T}^{\rm miss}}}$)
and for single $W$ boson production ($\sigma_{W}$), for the latter of which the branching
ratio ($\Gamma$=0.24) for leptonic $W$ decay is taken into account~\cite{h1wpol}.
For the isolated lepton cross section, $\Gamma=1$.
The cross sections are measured in the phase space
$5^{\circ}~<~\theta_{\ell}~<~140^{\circ}$, $P_{T}^{\ell}>$~10~GeV,
$P_{T}^{\rm miss}>$~12~GeV, and where the lepton is isolated from any jet by at least
one unit in $\eta-\phi$.


Cross sections are calculated using the formula:
\begin{equation}
\sigma = \frac{N_{data} - N^{MC}_{bkd}}{\mathcal{L}\,\Gamma\mathcal{A}} \qquad \textrm{with} \qquad
\mathcal{A} = \frac{N^{MC}_{rec}}{N^{MC}_{gen}},
\label{eqn:xs}
\end{equation}
where $N_{data}$ is the number of data events, $N^{MC}_{bkd}$ is the Monte Carlo (MC)
estimate of number of background events and $\mathcal{L}$ is the total data luminosity.
$\mathcal{A}$ is the acceptance, where $N^{MC}_{rec}$ and $N^{MC}_{gen}$ are
the number of reconstructed and generated events in the signal MC, respectively.
SM signal processes are described in section \ref{sec:isointro}. It should be noted that the
small SM contribution to $N^{MC}_{rec}$ from $Z^{0}$ production is signal for
$\sigma_{\ensuremath{\ell+{P}_{T}^{\rm miss}}}$, whereas it is considered background for $\sigma_{W}$. 


The results are shown in table \ref{tab:crosssections} with statistical (stat)
and systematic (sys) uncertainties and are found to be in good agreement with the SM predictions.

\begin{table*}[t]
  \begin{center}
  \caption{The measured H1 cross section for events with an isolated lepton and large
	missing transverse momentum ($\sigma_{\ensuremath{\ell+{P}_{T}^{\rm miss}}}$) and for
	single $W$ production ($\sigma_{W}$) with statistical (stat.) and systematic (sys.)
	uncertainties. The results are compared to the SM prediction with the associated
	theory uncertainty (th.).}
  \vspace{0.2cm}
  \label{tab:crosssections}
    \begin{tabular}{l c c }
    \hline
    {} & {H1 HERA I+II Data} & {SM} \\
    \hline
    {$\sigma_{\ensuremath{\ell+{P}_{T}^{\rm{miss}}}}$} &
    {0.24 $\pm$ 0.05 (stat.) $\pm$ 0.05 (sys.)} &
    {0.26 $\pm$ 0.04 (th.)} \\
    {$\sigma_{W}$} &
    {1.23 $\pm$ 0.25 (stat.) $\pm$ 0.22 (sys.)} &
    {1.31 $\pm$ 0.20 (th.)} \\
    \hline
    \end{tabular}
  \end{center}
\end{table*}


\section{Events with an Isolated Tau and $P_{T}^{\rm miss}$}
\label{sec:tau}

\begin{figure}[t]
  \includegraphics[width=.47\textwidth]{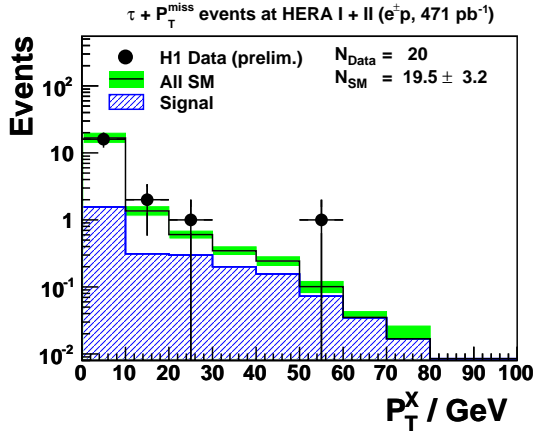}
  \vspace{-1cm}
  \caption{The hadronic transverse momentum distribution of
	$\tau + P_T^{\rm miss}$ events in the H1 $e^{\pm}p$ HERA I+II data.
	The data (points) are compared to the SM expectation (open histogram).
	The signal component of the SM expectation is given by the
	striped histogram. $\rm N_{Data}$ is the total number of data
	events observed and $\rm N_{SM}$ is the total SM expectation. The
	total uncertainty on the SM expectation is given by the shaded band.}
  \label{fig:tau}
\end{figure}

The H1 Collaboration has also performed a search for events with an isolated tau
lepton and large missing transverse momentum, using the full HERA~I+II $e^{\pm}p$
data and the hadronic one--prong tau decay mode~\cite{h1isotaunew}.
This search is complementary to the electron and muon searches described above,
and provides a test of lepton universality.
In addition, some BSM scenarios favour the third lepton generation, which could
lead to an enhancement from the subsequent leptonic decay of the tau lepton.


The event selection is based on that used in~\cite{tauheraI}, with improvements
to the tau identification algorithm, in particular to the track
isolation~\cite{h1isotaunew}.
The event signature is a narrow, low track multiplicity jet (tau--jet) in
coincidence with missing transverse momentum.


The hadronic transverse momentum distribution of the final sample is shown in
figure \ref{fig:tau}, where 20 events are observed in the data compared to a SM
prediction of 19.5~$\pm$~3.2.
The latter is dominated by charged current events and the signal
purity is observed to be much lower than the electron muon channels, at around 14\%.
For $P_{T}^{X}>$~25~GeV one event is selected in the data, compared to a SM
prediction of $0.99~\pm~0.13$. 
%


\section{Measurement of the $W$ Boson Polarisation Fractions}
\label{sec:pol}

The H1 measurement of the $W$ boson polarisation fractions at HERA makes use of the \cosths\
distributions in the decay $W\rightarrow e/\mu+\nu$, where $\theta^{*}$ is defined as the
angle between the $W$ boson momentum in the lab frame and the charged decay lepton in the $W$
boson rest frame.
The left handed $F_{-}$, longitudinal $F_{0}$ and right handed $F_{+}$ polarisation
fractions are constrained by the relation $F_{+} \equiv 1 - F_{-} - F_{0}$.
The \cosths\ distributions for $W^{+}$ bosons are given~\cite{wpoltheory} by:
\begin{eqnarray}
\frac{dN}{d\cos\,\theta^{*}} 
&\propto& \left( 1 - F_{-} - F_{0} \right) \cdot \frac{3}{8} \left( 1 + \cos\,\theta^{*}\right)^{2} \nonumber\\
&+&       F_{0} \cdot \frac{3}{4} \left( 1 - \cos^{2}\theta^{*}\right) \nonumber\\
&+&       F_{-} \cdot \frac{3}{8} \left( 1 - \cos\,\theta^{*}\right)^{2}.
\label{eqn:polmodel}
\end{eqnarray}
For $W^{-}$ bosons the \cosths\ distributions have opposite sign. To allow
the combination of the different $W$ boson charges, the value of \cosths\ is multiplied
by the sign of the lepton charge $q_{\ell} = \pm 1$.


Starting from the H1 event sample described in section \ref{sec:sep}, events are
selected for which a $W$ boson can be successfully reconstructed,
based on the method described in~\cite{wpolrecon}.
Additional selection criteria are applied to ensure a reliable charge measurement,
so that the resulting charge misidentification is well below $1\%$~\cite{h1wpol}.
Electron and muon events originating from tau decays of $W$ bosons are considered
background in this analysis,~since for these events the \cosths\ distributions
are not expected to be described by equation~\ref{eqn:polmodel}.
The final event sample consists of 21 electron events and 10 muon events with an
overall signal purity of 83\%.

\begin{figure}[p]
  \includegraphics[width=.47\textwidth]{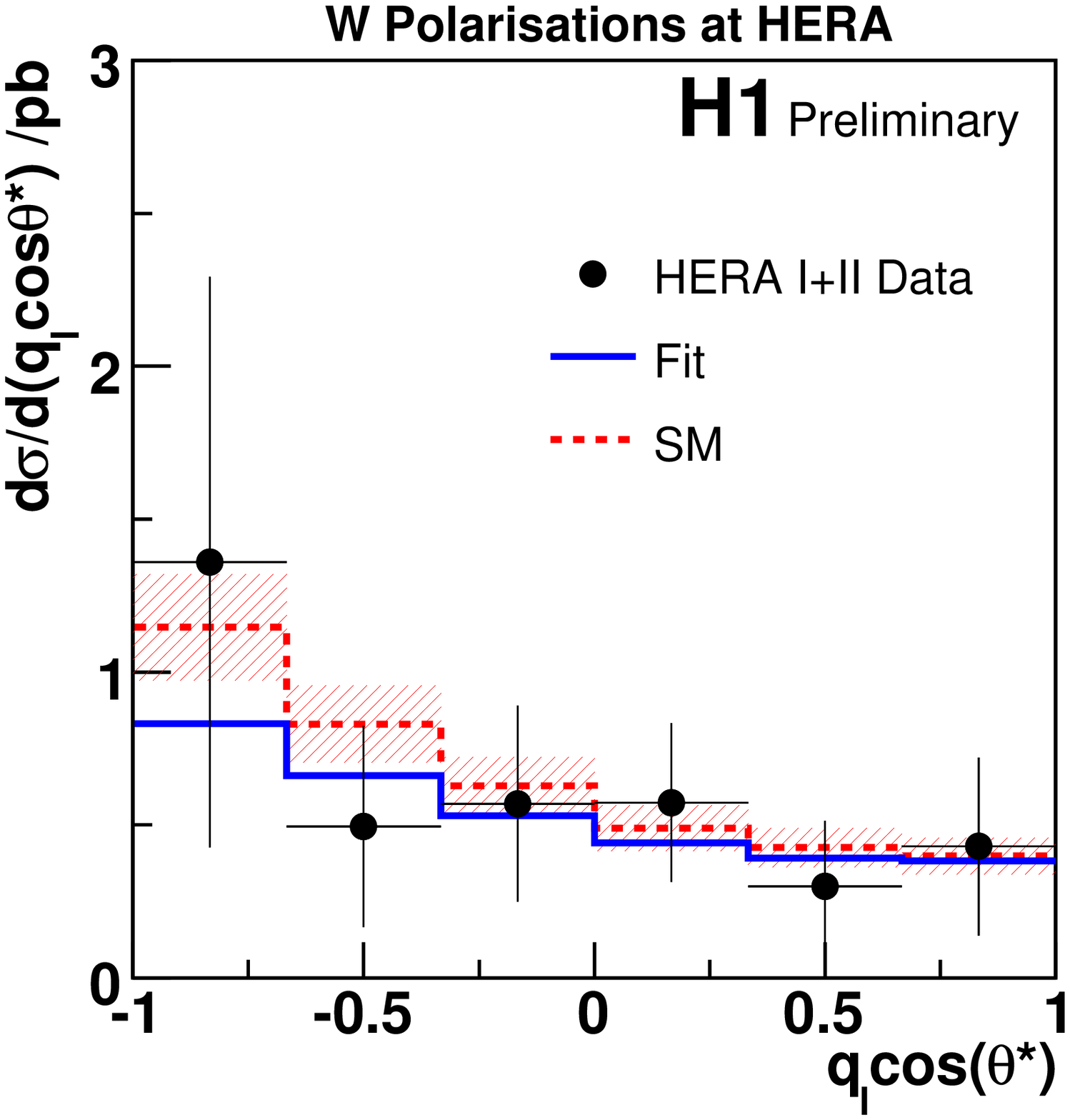}
  \includegraphics[width=.47\textwidth]{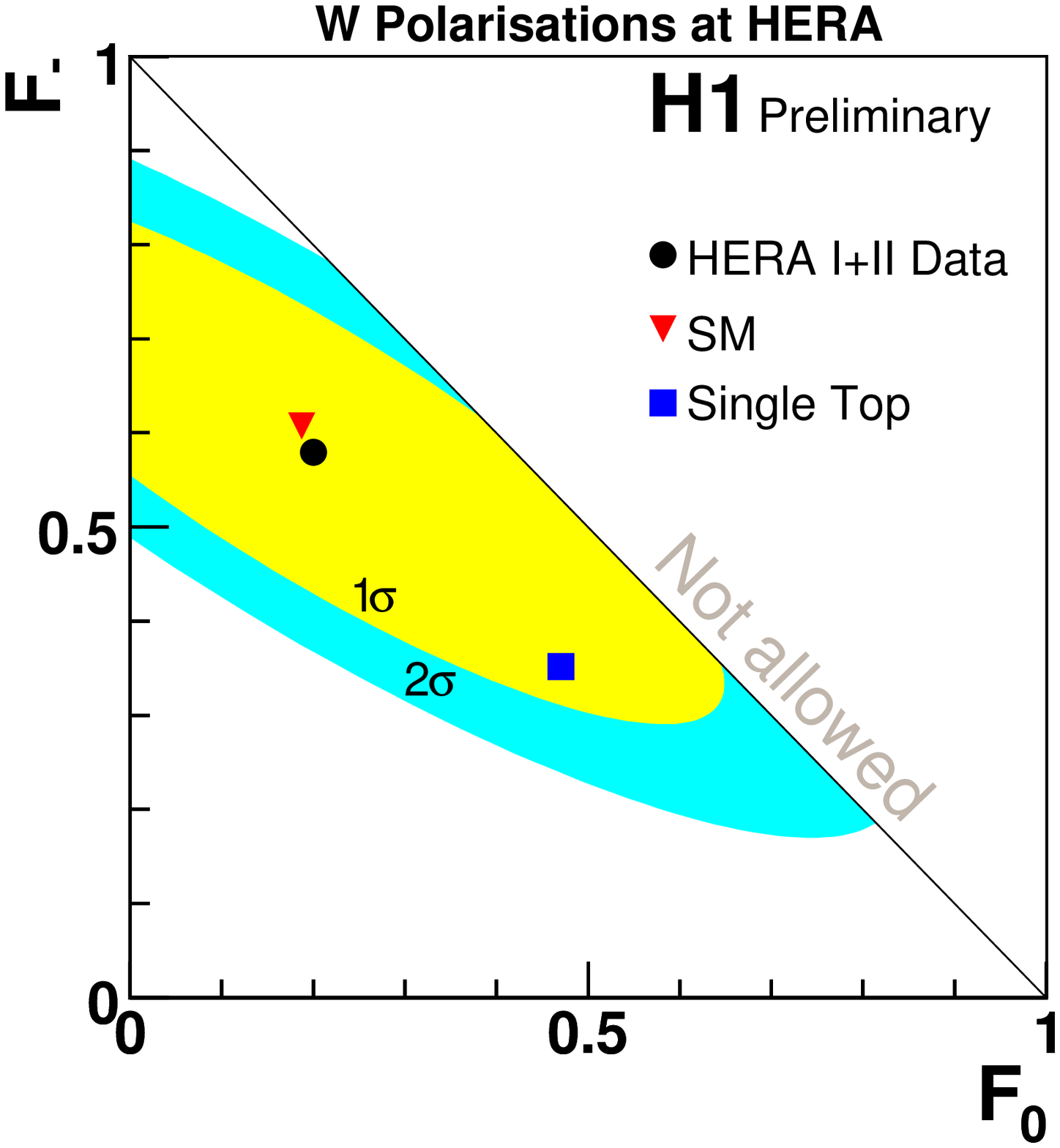}
  \vspace{-1cm}
  \caption{Top: A simultaneous fit (solid histogram) of $F_{-}$ and $F_{0}$ 
	to the measured differential cross section
	(points), where the error bars denote the statistical uncertainty only.
	The SM prediction (dashed histogram) is shown with a 15\% theoretical
	systematic uncertainty (hatched area). Bottom: The fit result for the
	simultaneously extracted $W$ boson polarisation
	fractions $F_{-}$ and $F_{0}$ (point) with 1 and 2$\sigma$ CL contours. The predictions for the
	SM prediction (triangle) and anomalous single top production via FCNC (square)
	are also shown.}
  \label{fig:pol}
\end{figure}

The measured \qcosths\ distribution, corrected for acceptance and detector effects,
is shown in figure \ref{fig:pol} (top), compared to the SM prediction.
The cross section fit to the model defined in equation~\ref{eqn:polmodel} is also shown.
In the fit, the optimal values for $F_{-}$ and $F_{0}$ are simultaneously extracted, the
result of which is shown in figure~\ref{fig:pol} (bottom).
The measured $W$ boson polarisation fractions are found to be in good agreement with the SM
prediction and compatible with anomalous single top production via FCNC within 1$\sigma$
confidence level (CL).


Values of $F_{-}$ and $F_{0}$ are also extracted in fits where one parameter is fixed
to its SM value. The results are presented in table~\ref{tab:pol} and show good agreement
with the SM.
The quoted systematic uncertainties are propagated from the differential cross section calculations.

\begin{table*}[t]
  \begin{center}
  \caption{One parameter fit results by H1 of the $W$ polarisation fractions $F_{-}$
	and $F_{0}$ with statistical (stat.) and sytematic (sys.) uncertainties.
	The central values are obtained by fixing one parameter to the SM prediction and fitting
	the other. The SM prediction is obtained from a two parameter fit to the SM
	$q_{\ell} \cos \theta^{*}$ distribution, where the quoted uncertainty is statistical only.}
  \vspace{0.2cm}
  \label{tab:pol}
    \begin{tabular}{l c c }
    \hline
    {} & {H1 HERA I+II Data} & {SM} \\
    \hline
    {$F_{-}$} &
    {0.58 $\pm$ 0.15 (stat.) $\pm$ 0.12 (sys.)} &
    {0.61 $\pm$ 0.01 (stat.)} \\
    {$F_{0}$} &
    {0.15 $\pm$ 0.21 (stat.) $\pm$ 0.09 (sys.)} &
    {0.19 $\pm$ 0.01 (stat.)} \\
    \hline
    \end{tabular}
  \end{center}
\end{table*}


\section{Search for Anomalous Single Top Quark Production}

The production of single top quarks is kinematically possible in $ep$ collisions
at HERA due to the large centre of mass energy of up to $\sqrt s=320$~GeV.
The dominant SM process for single top production at HERA is the charged current reaction
$e^+p \rightarrow \bar{\nu} t \bar{b} X$ ($e^-p \rightarrow \nu \bar{t} b X$), which
has a negligible cross section of less than~1~fb.
However, in several extensions of the SM the top quark is predicted to undergo
Flavour Changing Neutral Current (FCNC) interactions, which could lead to a sizeable
anomalous single top production cross section at HERA~\cite{toptheory}.


Such events are of interest as the signature of the top decay to $b$ and $W$ with
subsequent leptonic decay matches that of the isolated lepton events discussed in
section \ref{sec:isointro}.
In particular, the hadronic final state from the fragmentation of the $b$ quark
would exhibit high $P_T$, and thus this process could provide an explanation of the
data excess observed at high $P_{T}^{X}$ by the H1 experiment.\footnote{Although it should
be noted that single top production cannot explain the observed difference
between the H1 $e^{+}p$ and $e^{-}p$ data.}


Anomalous single top production is described by an effective Lagrangian where the interaction 
of a top with $u$-type quarks via a photon is described by a magnetic coupling $\kappa_{tU\gamma}$.
The contribution from the charm quark is expected to be small at the large proton
longitudinal momentum fraction $x$ needed to produce a top quark, and is
therefore neglected ($\kappa_{tc\gamma} \equiv 0$).
The vector couplings to a $Z^0$ boson $v_{tUZ}$ are supressed at HERA
due to the large $Z^{0}$ mass, and are also neglected in the present
H1 analysis ($v_{tUZ} \equiv 0$).


The H1 search for single top production is based on the full sample of events selected
in the HERA~I+II data described in section \ref{sec:sep}, so that the present analysis
investigates the leptonic $W$ decay channels to electrons and muons. 
The first step in the analysis forms a {\it top preselection} of this event sample, by demanding
good top quark reconstruction and lepton charge compatible with single top production~\cite{h1topnew}.
Three observables are investigated in this preselection, namely: the transverse momentum of the
reconstructed $b$ quark candidate, the reconstructed top mass, and the $W$ decay angle~calculated
as the angle between the lepton momentum in the $W$ rest frame and the $W$ direction in
the top quark rest frame.
The observed data distributions of these quantities agree well with the SM expectation
within the uncertainties and no evidence for single top production is observed.


The observables are then combined into a multi-variate discriminator, which is trained using
a single top MC as the signal model and a $W$ boson MC as the background model. 
The resulting discriminator output for the electron and muon channels is found to provide
good separation between $W$ and top MC events.


Limits on the signal cross section are extracted from the discriminator spectra using a
maximum likelihood method~\cite{h1topnew}.
Likelihood functions are calculated for the electron and muon channel separately.
%
%
%
An upper bound on the cross section of $\sigma_{ep \rightarrow e t X}<$~0.16~pb at 95\% CL
is found, which is translated into an upper bound on the coupling $\kappa_{tu\gamma}<$~0.14.

\begin{figure}[h]
  \includegraphics[width=.47\textwidth]{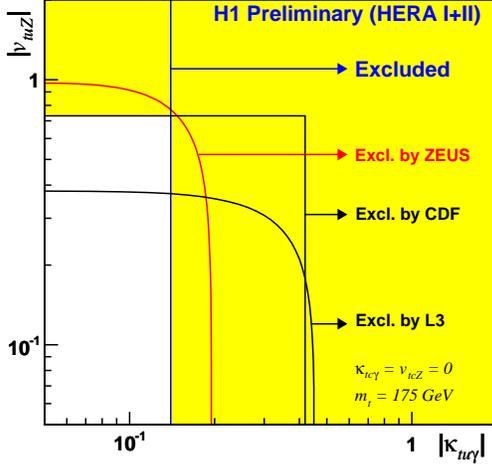}
  \vspace{-1cm}
  \caption{Exclusion limits at 95\% CL on  the anomalous $\kappa_{tu\gamma}$
	and $v_{tuZ}$ couplings obtained at HERA (H1~\cite{h1topnew} and
	ZEUS~\cite{zeustop} experiments), LEP (L3 experiment~\cite{l3top})
	and at the TeVatron (CDF experiment~\cite{cdftop}). Anomalous couplings
	of the charm quark are neglected $\kappa_{tc\gamma}$.
	Limits are shown assuming a top mass $m_t=175$~GeV.}
  \label{fig:toplimit}
\end{figure}

Figure \ref{fig:toplimit} shows existing limits on the anomalous couplings $\kappa_{tu\gamma}$
and $v_{tuZ}$.
The top mass is assumed to be $m_t = 175$~GeV in order to compare with previous results.
The new preliminary H1 result presented here extends the bound on $\kappa_{tu\gamma}$ into a region so
far uncovered by current colliders. 
Also shown in figure \ref{fig:toplimit} are results from the L3 experiment at LEP~\cite{l3top},
the CDF experiment at the Tevatron~\cite{cdftop} and results from the ZEUS experiment using
HERA~I data~\cite{zeustop}.
A new result from CDF~\cite{cdftopnew}, not shown in figure~\ref{fig:toplimit}, derives a limit
on the branching ratio $B(t \rightarrow Zq)$ of 3.7\%, which translates as an upper limit on
the anomalous vector coupling of $v_{tuZ}\lesssim$~0.2 and is the strictest limit to date.

\section{Summary}

Searches for events containing isolated leptons and missing transverse momentum
produced in $e^\pm p$ collisions at HERA are presented, performed individually
and in a common phase space with the H1 and ZEUS detectors at HERA in the
period 1994--2007.
In the complete H1+ZEUS high energy data sample, luminosity 0.97~fb$^{-1}$, a total of 87
events are observed in the data, compared to a Standard Model (SM) prediction of
92.7~$\pm$~11.2.
At large hadronic transverse momentum $P_{T}^{X} >$~25~GeV in the $e^{+}p$ data,
luminosity 0.58~fb$^{-1}$, 23 data events are observed compared to a SM prediction of 14.6~$\pm$~1.9.
The observed excess is dominated by the H1 data.


Production cross section measurements of events containing isolated
leptons and missing transverse momentum and single $W$ production are
performed by H1, as well as a measurement of the $W$ polarisation fractions.
The H1 isolated lepton events are also examined in the context of a search for
single top production.
No clear signal is observed, and an upper limit on the anomalous top production
cross section of $\sigma_{ep\rightarrow etX} < 0.16$~pb is established
at the $95\%$ confidence level.


\end{document}